\documentclass[12pt]{article}

\usepackage{color}
\usepackage{array}
\usepackage{epsfig}
\usepackage{amssymb}
\usepackage{graphics,graphpap}
\usepackage{subfigure}
\usepackage{amsmath}

\newcommand{\bal}{\begin{align}}

\def\beq{\begin{equation}}
\def\eeq{\end{equation}}
\def\bea{\begin{eqnarray}}
\def\eea{\end{eqnarray}}
\def\nn{\nonumber}
\def\nl{\nonumber\\}
\def\sss{\scriptscriptstyle}
\def\roughly#1{\mathrel{\raise.3ex\hbox
{$#1$\kern-.75em\lower1ex\hbox{$\sim$}}}}

\def\lesssim{\mathrel{\hbox{\rlap{\hbox{\lower4pt\hbox{$\sim$}}}\hbox{$<$}}}}
\def\gtrsim{\mathrel{\hbox{\rlap{\hbox{\lower4pt\hbox{$\sim$}}}\hbox{$>$}}}}

\def\sla#1{\raise.15ex\hbox{$/$}\kern-.57em #1}
\def\bra#1{\left\langle #1\right|}
\def\ket#1{\left| #1\right\rangle}

\newcommand{\ba}{\begin{array}}
\newcommand{\ea}{\end{array}}

\def\bs{B_s^0}

\def\bsbar{{\bar B}_s^0}
\def\btod{{\bar b} \to {\bar d}}
\def\btos{{\bar b} \to {\bar s}}
\def\btoccs{{\bar b} \to {\bar c} c {\bar s}}
\def\btouus{{\bar b} \to {\bar u} u {\bar s}}
\def\btosqq{{\bar b} \to {\bar s} q {\bar q}}

\def\btosdd{{\bar b} \to {\bar s} d {\bar d}}

\def\Bsdecay{\bs\to J/\psi \phi}

\def\BsKK{\bs \to K^{*0} {\bar K}^{*0}}

\def \({\left(}
\def \){\right)}
\def \s{\sqrt{2}}
\def \st{\sqrt{3}}

\def \sx{\sqrt{6}}

\def \i{{\it i}}

\def \G{\Gamma}
\def \hf{\frac{1}{2}}
\def \ot{\frac{1}{3}}
\def \ox{\frac{1}{6}}
\def \on{\frac{1}{9}}
\def\bA{{\bar A}}
\def\Bsdecay{\bs\to J/\psi \phi}

\allowdisplaybreaks
\begin{document}

\begin{flushright}  
UMISS-HEP-2013-06\\
UdeM-GPP-TH-13-224 \\
\end{flushright}

\begin{center}
\bigskip
{\Large \bf \boldmath Searching for New Physics with \\ $\btos$
  $\bs \to V_1 V_2$ Penguin Decays} \\

\bigskip
{\large 
Bhubanjyoti Bhattacharya $^{a,}$\footnote{bhujyo@lps.umontreal.ca},
Alakabha Datta $^{b,}$\footnote{datta@phy.olemiss.edu}, \\
Murugeswaran Duraisamy $^{b,}$\footnote{duraism@phy.olemiss.edu}
and David London $^{a,}$\footnote{london@lps.umontreal.ca}}
\\
\end{center}

\begin{flushleft}
~~~~~~~~~~~$a$: {\it Physique des Particules, Universit\'e
de Montr\'eal,}\\
~~~~~~~~~~~~~~~{\it C.P. 6128, succ.\ centre-ville, Montr\'eal, QC,
Canada H3C 3J7}\\
~~~~~~~~~~~$b$: {\it Department of Physics and Astronomy, 108 Lewis Hall, }\\ 
~~~~~~~~~~~~~~~{\it University of Mississippi, Oxford, MS 38677-1848, USA}\\
\end{flushleft}
\begin{center} 
\vskip0.5cm
{\Large Abstract \\}
\vskip3truemm

\parbox[t]{\textwidth} {We present the most general (six-helicity)
  angular analysis of $\bs \to V_1 (\to P_1P'_1) V_2 (\to P_2P'_2)$
  ($V_i$ is a vector meson, and $P_i, P_i'$ are pseudoscalars). We
  focus on final states accessible to both $\bs$ and $\bsbar$ -- these
  are mainly $\btos$ penguin decays. We also derive the most general
  decay amplitude, and discuss the differences between it and that
  used by LHCb in its analysis of $\bs \to \phi \phi$. In the standard
  model, all CP violation is predicted to be small, so that the simple
  measurement of a sizeable CP-violating observable indicates the
  presence of new physics. A full fit to the data is not necessary. By
  determining which of the CP-violating observables are nonzero, one
  can learn about the structure of the underlying NP.  Finally, we
  apply the angular analysis to $\bs \to K^{*0} {\bar K}^{*0}$, and
  show that there are numerous CP-violating observables that remain in
  the untagged data sample.}

\end{center}

\bigskip
\leftline{PACS numbers: 11.30.Er, 12.15.Ji, 13.25.Hw, 14.40.Nd}
\bigskip

\baselineskip=14pt

\newpage

\section{Introduction}

$B\to V_1V_2$ decays ($V_i$ is a vector meson) are really three
separate decays, one for each polarization of the final state (one
longitudinal, two transverse). Here it is useful to use the linear
polarization basis, where one decomposes the decay amplitude into
components in which the polarizations of the final-state vector mesons
are either longitudinal ($A_0$), or transverse to their directions of
motion and parallel ($A_\|$) or perpendicular ($A_\perp$) to one
another. Many years ago it was shown that one can separate these three
helicities by performing an angular analysis of the decay
\cite{angdist}.

Recently, it was pointed out that, under certain circumstances,
modifications must be made to the angular analysis. In particular,
when a neutral vector meson is detected via its decay $V \to P P'$
($P, P'$ are pseudoscalars), there is usually a background coming from
the decay of a scalar resonance $S \to P P'$, or from the scalar
non-resonant $P P'$ production \cite{Swave}. As such, it is necessary
to add another (scalar) helicity to the angular analysis. The LHCb
Collaboration performed this addition in their studies of the decays
$\bs \to J/\psi \phi$ \cite{BsJpsiphiLHCb} and $\bs \to \phi \phi$
\cite{BsphiphiLHCb}. In both cases the $\phi$ is detected through its
decay to $K^+ K^-$, and there is a resonant ($f_0$) or non-resonant
scalar background. Thus, the angular analyses in
Refs.~\cite{BsJpsiphiLHCb} and \cite{BsphiphiLHCb} were performed with
four and five helicities, respectively.

However, in the experimental analysis of $\bs \to \phi \phi$
\cite{BsphiphiLHCb}, the most general $\bs \to \phi \phi$ amplitude
was not used\footnote{In the study of $\bs \to J/\psi \phi$
  \cite{BsJpsiphiLHCb}, the most general angular analysis was also not
  performed. Rather, simplifying assumptions were imposed. The
  importance of including the most general amplitude was stressed in
  Ref.~\cite{Datta:2009fk}, and in Ref.~\cite{pengpoll} it was pointed
out that the penguin pollution can be reduced if the assumptions are
not made, and the full angular analysis done.}. Rather,
  simplifications were made based on approximations that hold only
  within the standard model (SM).  This then implied that certain
  new-physics (NP) signals were absent from the angular analysis. But
  since the goal is to seek signals of NP, it does not make sense to
  do only a SM-based angular analysis\footnote{A first attempt at a
    theoretical analysis of $\bs \to \phi \phi$ with the general
    amplitude including NP was presented in
    Ref.~\cite{Datta:2012ky}.}.  Furthermore, not all the SM
  assumptions were physically well-motivated.  We must stress that the
  main result of Ref.~\cite{BsphiphiLHCb} -- that there is a potential
  disagreement with the predictions of the SM -- is not in
  question. Our point is simply that it was not sufficiently precise
  {\it what} this disagreement is, and what further NP signals are
  possible.

In addition, we were informed that LHCb is studying the decay $\bs \to
K^{*0}(892) {\bar K}^{*0}(892)$, and that each of these vector mesons
has a background coming from the scalar resonance $K^{*0}(1430)$
\cite{Bernardo}. It is therefore necessary to perform an angular
analysis that takes this background into account. In this case, as one
does not have identical particles in the final state (in contrast to
$\bs \to \phi \phi$), six helicities must be considered.

In light of all of this, we feel it is useful to present the most
general angular analysis of $\bs \to V_1 (\to P_1P'_1) V_2 (\to
P_2P'_2)$. We focus on final states accessible to both $\bs$ and
$\bsbar$, which are mainly $\btos$ penguin decays. Our analysis allows
for the presence of NP, and we discuss the possible NP signals. Given
the LHCb constraints on NP in $\bs$-$\bsbar$ mixing
\cite{BsJpsiphiLHCb}, there is little sensitivity to NP of this
type. However, $\bs \to V_1 V_2$ decays can probe NP in the decay. In
particular, since $\btos$ penguin decays are dominated by a single
contributing amplitude in the SM, all CP-violating observables are
predicted to be small. The observation of sizeable CP violation would
then be a smoking-gun signal of NP in the decay. In fact, although
experiments aim to search for NP via a complete fit to the data, such
a fit is not really necessary. A more direct way to detect NP is
simply to measure a large CP-violating observable. In addition, one
can get an idea about the structure of the underlying NP from the
pattern of the measurements (i.e., determining which of the
CP-violating observables are nonzero).

We apply these ideas to the decays $\bs \to \phi \phi$ and $\bs \to
K^{*0} {\bar K}^{*0}$. For $\bs \to \phi \phi$, we compare the
amplitude used in Ref.~\cite{BsphiphiLHCb} with the exact amplitude,
and examine how the differences can affect the fit.  For $\bs \to
K^{*0} {\bar K}^{*0}$, we use the full (six-helicity) angular analysis
to detail which CP-violating observables remain in the untagged data
sample. This decay is particularly interesting since, in addition to
triple products, certain direct and indirect CP asymmetries can be
observed in untagged decays.

In Sec.~2, we present the full six-helicity angular distrinution. We
address the question of new physics in Sec.~3. Here we point out that
the best way to search for NP is to measure CP-violating observables,
and we discuss the four types of observables. In Secs.~4 and 5, we
apply the formalism to the decays $\bs \to \phi \phi$ and $\bs \to
K^{*0} {\bar K}^{*0}$. We examine a particular model of NP in
Sec.~6. Here we show that different NP operators lead to different
patterns of nonzero CP-violating observables. We conclude in Sec.~7.

\section{General Angular Distribution}

We consider the decay $\bs \to V_1V_2$. As discussed in the
introduction, when either vector meson decays to two pseudoscalar
mesons, there is generally a background due to the (resonant or
non-resonant) scalar production of the two pseudoscalars. We therefore
focus on the decay $\bs \to V_1/S_1 (\to P_1P'_1) V_2/S_2 (\to
P_2P'_2)$, concentrating on final states to which both $\bs$ and
$\bsbar$ can decay. 

In general, there are 6 helicities: $h = VV$ (3), $VS$, $SV$, and $SS$,
each with a corresponding amplitude $A_h$. Thus, when the full
amplitude is squared, there are 21 terms. Due to $\bs$-$\bsbar$
mixing, the amplitude is time dependent.  The angular distribution can
be written
\bea
\frac{d^4\G(t)}{dt d\cos\theta_1 d\cos\theta_2 d\phi} &=& \frac{9}{8 \pi} \sum^{21}_{i=1} K_i(t) X_i(\theta_1,\theta_2,\phi) ~,
\eea
where $\theta_1, \theta_2$ and $\phi$ are the helicity angles: $\theta_1$
($\theta_2$) is the angle between the directions of motion of the
$P_1$ ($P_2$) in the $V_1$ ($V_2$) rest frame and the $V_1$ ($V_2$) in
the $B$ rest frame, and $\phi$ is the angle between the normals to the
planes defined by $P_1 P_1'$ and $P_2 P_2'$ in the $B$ rest frame. 

\subsection{\boldmath $t=0$}\label{sec:tiad}

Much can be learned by studying the behaviour at $t=0$. The individual
amplitudes are constructed as follows:
\bea
{\cal A}_{VV} &=& N~\sum\limits^1_{j = -1} A^{VV}_j Y^{-j}_1(\theta_1,-\phi) Y^j_1(\pi-\theta_2,
0)~, \nn \\
{\cal A}_{VS} &=& N~A^{VS}_0 Y^0_1(\theta_1, -\phi) Y^0_0(\pi-\theta_2, 0)~, \nn\\
{\cal A}_{SV} &=& N~A^{SV}_0 Y^0_0(\theta_1, -\phi) Y^0_1(\pi-\theta_2, 0)~, \nn\\
{\cal A}_{SS} &=& N~A^{SS}_0 Y^0_0(\theta_1, -\phi) Y^0_0(\pi-\theta_2, 0)~.
\eea
N is a normalization constant and the $Y^m_l$ are spherical
harmonics. Using the standard expressions for the $Y^m_l$, as well as
$A_\| = (A_+ + A_-)/\sqrt{2}$ and $A_\perp = (A_+ - A_-)/\sqrt{2}$, we
have
\bea
{\cal A}_{VV} + {\cal A}_{VS} + {\cal A}_{SV} + {\cal A}_{SS} = -\frac{3N}{4\pi}\(A_0\cos\theta_1\cos\theta_2
- \frac{A_S}{3} - \frac{A_{VS}} {\st}\cos\theta_1 + \frac{A_{SV}}{\st}\cos\theta_2\right.\nn\\
\left. +~\frac{A_\|}{\s}\sin\theta_1\sin\theta_2\cos\phi + \i\frac{A_\perp}{\s}\sin\theta_1\sin
\theta_2\sin\phi\) ~. \label{eq:5}
\eea

It is convenient to choose a different notation for the $VS$ modes. We
introduce the following amplitude coefficients:
\beq
A^{(VS)}_+ \equiv \frac{A_{VS} + A_{SV}}{\s}~,~~~A^{(VS)}_- \equiv \frac{A_{VS} - A_{SV}}{\s}~.
\label{CPVS}
\eeq
Using this notation, Eq.\ (\ref{eq:5}) can be rewritten as follows:
\bea
{\cal A}_{VV} + {\cal A}_{VS} + {\cal A}_{SV} + {\cal A}_{SS} = -\frac{3N}{4\pi}\(A_0\cos\theta_1\cos\theta_2
- \frac{A_S}{3} - \frac{A^{(VS)}_+}{\sx}(\cos\theta_1 - \cos\theta_2) \right.\nn\\
\left. - \frac{A^{(VS)}_-}{\sx}(\cos\theta_1+\cos\theta_2) + \frac{A_\|}{\s}\sin\theta_1\sin\theta_2\cos\phi
+ \i\frac{A_\perp}{\s}\sin\theta_1\sin\theta_2\sin\phi\) ~.
\label{eq:6}
\eea

We can now construct the $t=0$ differential decay rate:
\bea
\frac{d^4\G}{dt d\cos\theta_1 d\cos\theta_2 d\phi} &=& \frac{9}{8\pi}|{\cal A}_{VV} + {\cal A}_{VS} +
{\cal A}_{SV} + {\cal A}_{SS}|^2 \nn \\
&=&\frac{9}{8\pi}\sum\limits^{21}_{n = 1} K_n X_n(\theta_1,\theta_2,\phi) ~.
\label{eq:b}
\eea
In Table \ref{tab:KX} we list the individual $K$'s and $X$'s. The
normalization constant $N$ has been chosen such that the integration
of Eq.\ (\ref{eq:b}) over the entire phase space gives
\beq
\frac{d\G}{dt} = \sum_h |A_h|^2 ~~=~~|A_0|^2 + |A_\||^2 + |A_\perp|^2 + |A^{(VS)}_+|^2
+ |A^{(VS)}_-|^2 + |A_S|^2 ~.
\eeq

The various angular functions can be isolated by performing asymmetric
integrals over the three angles \cite{bsVV}. For example, consider
$n=5$: $X_5 = -(1/2\s) \sin2\theta_1\sin2\theta_2\sin\phi$. If one
integrates over $0\le\phi\le\pi$ with a $+$ sign, and $\pi\le\phi\le
2\pi$ with a $-$ sign, one eliminates all the other $X_n$ except those
proportional to $\sin\phi$ ($n=12,17,19$). These can be eliminated by
integrating asymmetrically over each of $\theta_{1,2}$: $0\le
\theta_{1,2}\le\pi/2$ with a $+$ sign, and $\pi/2 \le
\theta_{1,2}\le\pi$ with a $-$ sign. The other $X_n$ can be isolated
similarly. The one exception involves $n=7,13,20$. The difference $X_7
- X_{13}$ is proportional to $\cos\theta_1\cos\theta_2$, as is
$X_{20}$. These two can therefore not be differentiated. However,
apart from this lone exception, all the $X_n$ can be isolated
experimentally.

\begin{table}[htb]
\caption{Individual $K$'s and $X$'s listed in Eq.\ (\ref{eq:b})
\label{tab:KX}}
\centering
\begin{tabular}{|c|c|c|} \hline \hline
 $n$ & $K_n$ & $X_n$ \\ \hline
 1 & $|A_0|^2$     & $\cos^2\theta_1\cos^2\theta_2$     \\
 2 & $|A_\||^2$    & $\hf\sin^2\theta_1\sin^2\theta_2\cos^2\phi$ \\
 3 & $|A_\perp|^2$ & $\hf\sin^2\theta_1\sin^2\theta_2\sin^2\phi$ \\
 4 & ${\rm Re}[A_\|A^*_0]$& $\frac{1}{2\s}\sin2\theta_1\sin2\theta_2\cos\phi$ \\
 5 & ${\rm Im}[A_\perp A^*_0]$& $-\frac{1}{2\s}\sin2\theta_1\sin2\theta_2\sin\phi$ \\
 6 & ${\rm Im}[A_\perp A^*_\|]$& $-\hf\sin^2\theta_1\sin^2\theta_2\sin2\phi$ \\ \hline
 7 & $|A^{(VS)}_+|^2$  & $\ox(\cos\theta_1 - \cos\theta_2)^2$  \\
 8 & ${\rm Re}[A^{(VS)}_+A^*_S]$& $\frac{\s}{3\st}(\cos\theta_1 - \cos\theta_2)$ \\
 9 & ${\rm Re}[A^{(VS)}_+A^{(VS)*}_-]$& $\ot(\cos^2\theta_1 - \cos^2\theta_2)$ \\
10 & ${\rm Re}[A^{(VS)}_+A^*_0]$& $-\frac{\s}{\st}\cos\theta_1\cos\theta_2(\cos\theta_1 - \cos\theta_2)$ \\
11 & ${\rm Re}[A^{(VS)}_+A^*_\|]$& $-\frac{1}{\st}\sin\theta_1\sin\theta_2\cos\phi(\cos\theta_1 - \cos\theta_2)$ \\
12 & ${\rm Im}[A_\perp A^{(VS)*}_+]$& $\frac{1}{\st}\sin\theta_1\sin\theta_2\sin\phi(\cos\theta_1 - \cos\theta_2)$ \\ \hline
13 & $|A^{(VS)}_-|^2$  & $\ox(\cos\theta_1 + \cos\theta_2)^2$  \\
14 & ${\rm Re}[A^{(VS)}_- A^*_S]$& $\frac{\s}{3\st}(\cos\theta_1 + \cos\theta_2)$ \\
15 & ${\rm Re}[A^{(VS)}_-A^*_0]$& $-\frac{\s}{\st}\cos\theta_1\cos\theta_2(\cos\theta_1 + \cos\theta_2)$ \\
16 & ${\rm Re}[A^{(VS)}_-A^*_\|]$& $-\frac{1}{\st}\sin\theta_1\sin\theta_2\cos\phi(\cos\theta_1 + \cos\theta_2)$ \\
17 & ${\rm Im}[A_\perp A^{(VS)*}_-]$& $\frac{1}{\st}\sin\theta_1\sin\theta_2\sin\phi(\cos\theta_1 + \cos\theta_2)$ \\ \hline
18 & ${\rm Re}[A_S A^*_\|]$& $-\frac{\s}{3}\sin\theta_1\sin\theta_2\cos\phi$ \\
19 & ${\rm Im}[A_\perp A^*_S]$& $\frac{\s}{3}\sin\theta_1\sin\theta_2\sin\phi$ \\
20 & ${\rm Re}[A_SA^*_0]$& $-\frac{2}{3}\cos\theta_1\cos\theta_2$ \\
21 & $|A_S|^2$     & $\on$  \\\hline\hline
\end{tabular}
\end{table}

\subsection{Time-dependent angular distribution}
\label{sec:tdad}

Due to $B^0_s$-${\bar B}^0_s$ mixing, the time evolution of the states
$\ket{B^0_s (t)}$ and $\ket{{\bar B}^0_s(t)}$ can be described by the
relations
\bea
\label{Timeev:Bst}
\ket{B^0_s (t)} &=& g_+(t) \,\ket{B^0_s}  + \frac{q}{p}\, g_-(t) \, \ket{{\bar B}_s^0}   ~, \nl
\ket{{\bar B}_s^0 (t)} &=& \frac{p}{q}\, g_-(t) \,\ket{B_s^0}  + \, g_+(t) \, \ket{{\bar B}_s^0}   ~,
\eea
where $q/p = (V_{tb}^* V_{ts})/(V_{tb} V_{ts}^*) \equiv e^{-i\phi_M}$.
(Here we follow the notation of LHCb: $\phi_M$ is the theoretical phase
of $B^0_s$-${\bar B}^0_s$ mixing, while $\phi_s$ is its
experimentally-measured value.) In the above, we have
\bea
g_+(t) &=& \frac12 \Big(e^{-(i M_{L} + \Gamma_{L}/2)t} \,+ e^{-(i M_{H} + \Gamma_{H}/2)t}\Big)  ~, \nl
g_-(t) &=& \frac12  \Big(e^{-(i M_{L} + \Gamma_{L}/2)t} \,- e^{-(i M_{H} + \Gamma_{H}/2)t}\Big) ~,
\eea
where $L$ and $H$ indicate the light and heavy states, respectively.
The average mass and width are $m = (M_H + M_L)/2$ and $\Gamma =
(\Gamma_L + \Gamma_H)/2$, while the mass and width differences of the
$\bs$-meson eigenstates are defined as $\Delta m \equiv M_H - M_L$ and
$\Delta \Gamma \equiv \Gamma_L - \Gamma_H$.  $\Delta m$ is positive by
definition. For $\bs$ mesons, $\Delta \Gamma_s$ is reasonably large,
and is positive in our convention.

Now, the time dependence of the transversity amplitudes $A_h$ is due
to $B^0_s$-${\bar B}^0_s$ mixing. Their precise form depends
on the specific final state. The different helicities of the $VV$ and
the $SS$ final states are all CP eigenstates. On the other hand, the
$VS$ and $SV$ states are not CP eigenstates. However, their linear
combinations, defined as
$|\pm\rangle_{VS}\equiv(|VS\rangle\pm|SV\rangle )/\s$, {\it are} CP
eigenstates. Working only with CP eigenstates, the time-dependent
amplitudes are given as \cite{DF}
\bea
\label{Tdep:TransAmp}
A_h(t) &=&  \langle f | H_W|B^0_s (t) \rangle_h = \Big[ g_+(t) A_h+ \eta_h ~ q/p ~g_-(t)~ \bar{A}_h\Big] ~, \nl
\bar{A}_h(t) &=&  \langle f |H_W| \bar{B}^0_s (t) \rangle_h = \Big[ p/q~g_-(t) A_h+ \eta_h ~ g_+(t)~ \bar{A}_h\Big] ~,
\eea
where $A_h = \langle f |H_W|B^0_s \rangle_h$, $\bar{A}_h = \langle f
|H_W|\bar{B}^0_s \rangle_h$, and $\langle \bar{f} | = \eta_h \langle f |$,
with $\eta_h = +1$ for $h = 0, \|, VS_-, SS$ and $\eta_h = -1$ for $h
= VS_+, \perp$. These values for the CP eigenvalue $\eta_h$ can be
understood in terms of the total angular momentum of the final state.
States with $l=0$ ($SS$, $VS_-$, a combination of $0, \|$) and $l=2$
(another combination of $0, \|$) are CP even, while those with $l=1$
($VS_+$, $\perp$) are CP odd.  It is also important to point out the
CP properties of the helicity amplitudes $A_h$ and $\bA_h$:
\bea
CP~A_h~=~CP~\langle f|H_W|B^0_s \rangle_h~=~\langle\bar{f}|H_W|\bar{B}^0_s\rangle_h~=~\eta_h\langle f|\bar{B}^0_s\rangle_h
~=~\eta_h\bA_h~, \nl
CP~\bA_h~=~CP~\langle f|H_W|\bar{B}^0_s \rangle_h~=~\langle\bar{f}| H_W|B^0_s\rangle_h~=~\eta_h\langle f| B^0_s\rangle_h
~=~\eta_hA_h~.
\eea
Thus, in order to go from the $\bs$ decay to the $\bsbar$ decay, one
simply needs to switch $A_h \leftrightarrow \eta_h\bA_h$.

It is useful to discuss the origin of the $\eta_h$ factors in
Eq.~(\ref{Tdep:TransAmp}) since, naively, such factors are not
present. This is understood most easily by considering $B^0_s \to
K^{*0} \bar{K}^{*0}$.  As noted above, in the decay $B^0_s \to V_1
V_2$, the helicity angles are defined with respect to the momenta of
$V_1$ and $V_2$. In $B^0_s \to K^{*0} \bar{K}^{*0}$, $V_1 = K^{*0}$
and $V_2 = \bar{K}^{*0}$. On the other hand, in the CP-conjugate decay
${\bar B}^0_s \to \bar{K}^{*0} K^{*0}$ we have $V_1 = \bar{K}^{*0}$
and $V_2 = K^{*0}$. That is, we have $V_1 \leftrightarrow V_2$
compared to the $B^0_s$ decay. The effect of this on the helicity
angles is to change $\theta_1 \leftrightarrow \theta_2$ and $\phi \to
-\phi$. Looking at Table \ref{tab:KX}, we see that the $X_n$ change
sign for $n =$ 5, 6, 8-12, 19, i.e., when only one CP-odd state is
involved in the $K_n$. This sign change can be transferred to the
$A_h(t)$ by adding the $\eta_h$ factors in
Eq.~(\ref{Tdep:TransAmp}). That is, by using this definition for the
$A_h(t)$, the angular functions are the same for $B^0_s$ and ${\bar
  B}^0_s$ decays, which makes it easy to compute what is measured in
untagged samples.

The expressions for the time-dependent wave functions for the various
terms are
\bea\label{eq:tdep}
|A_i(t)|^2 & = & \frac12 e^{ - \Gamma t} \left[
\left(|A_i|^2 + |{\bar A}_i|^2 \right) \cosh{(\Delta \Gamma/2) t} \right. \nl
&& \hskip0.7truecm -~2~\eta_i~{\rm Re} \left( A^*_i \bA_i e^{-\i\phi_M} \right) \sinh{(\Delta\Gamma/2) t} \nl
&& \hskip0.7truecm  +~\left(|A_i|^2 - |{\bar A}_i|^2 \right) \cos{ \Delta m t} \nl
&& \hskip0.7truecm \left. -~2~\eta_i~{\rm Im} \left( A^*_i \bA_i e^{-\i\phi_M} \right) \sin{ \Delta m t}\right]~, \nl
{\rm Im} ( A_\perp(t) A_j^*(t) ) & = & \frac12 e^{ - \Gamma t} \left[
{\rm Im} \left( A_\perp A_j^* - \eta_j \bar{A}_\perp \bar{A}_j^* \right) \cosh{(\Delta \Gamma/2) t} \right. \nl
&& \hskip0.7truecm +~{\rm Im} \left[(\bar{A}_\perp A_j^* + \eta_j A^*_\perp \bar{A}_j)e^{-\i\phi_M}\right]
\sinh{(\Delta \Gamma/2)t} \nl
&& \hskip0.7truecm +~ {\rm Im} \left( A_\perp A_j^* + \eta_j \bar{A}_\perp \bar{A}_j^* \right) \cos{ \Delta m t} \nl
&& \hskip0.7truecm \left. -~{\rm Re} \left[(\bar{A}_\perp A_j^* + \eta_j A^*_\perp \bar{A}_j)e^{-\i\phi_M}\right]
\sin{ \Delta m t} \right] ~, \nl
{\rm Re} ( A_k(t) A_l^*(t) ) & = & \frac12 e^{ - \Gamma t} \left[
{\rm Re}\left(A_k A_l^* + \eta_k \eta_l \bar{A}_k \bar{A}_l^*\right) \cosh{(\Delta \Gamma/2) t} \right. \nl
&& \hskip0.7truecm +~{\rm Re}\left(A_k A_l^* - \eta_k \eta_l \bar{A}_k \bar{A}_l^* \right) \cos{ \Delta m t} \nl
&& \hskip0.7truecm -~{\rm Re} \left[ (\eta_k \bA_k A_l^* + \eta_l A^*_k \bA_l) e^{-\i\phi_M}\right] \sinh{(\Delta \Gamma/2) t} \nl
&& \hskip0.7truecm \left. -~{\rm Im} \left[ (\eta_k \bA_k A_l^* + \eta_l A^*_k \bA_l) e^{-\i\phi_M}\right]
 \sin{ \Delta m t} \right] ~.
\eea

Using the above, it is possible to write down the time dependence of
the functions $K_n$ listed in Table \ref{tab:KX}. In general, we have
\bea\label{eq:gentd}
K_n(t) = \frac12 e^{ - \Gamma t} \left[a_n \cosh{(\Delta \Gamma/2) t} + b_n \sinh{(\Delta\Gamma/2) t}
+ c_n \cos{\Delta m t} + d_n \sin{\Delta m t} \right]~,
\eea
where the individual functions $a_n, b_n, c_n$, and $d_n$ for $n = 1,
\dots, 21$ are time independent. In Tables \ref{tab:acs} and
\ref{tab:bds} we present the forms of the coefficients
$a_n$-$d_n$. These are exact and hold even in the presence of NP.
Note that not all of the $a_n$-$d_n$ are independent. There are 23
unknown parameters -- 12 magnitudes of $A_h$ and $\bA_h$, and 11
relative phases ($\phi_M$ can be absorbed into the phases of the $A_h$
and $\bA_h$) -- while there are 84 different $a_n$-$d_n$.  There are
therefore many relations among the $a_n$-$d_n$.

\begin{table}[htb]
\caption{$a_n$'s and $c_n$'s as defined in Eq.\ (\ref{eq:gentd})
\label{tab:acs}}
\centering
\begin{tabular}{|c|c|c|c|} \hline \hline
 $n$ & $K_n(t)$ & $a_n$ & $c_n$ \\ \hline
 1 & $|A_0(t)|^2$ & $|A_0|^2 + |\bA_0|^2$& $|A_0|^2 - |\bA_0|^2$  \\
 2 & $|A_\|(t)|^2$ & $|A_\||^2 + |\bA_\||^2$& $|A_\||^2 - |\bA_\||^2$ \\
 3 & $|A_\perp(t)|^2$ & $|A_\perp|^2 + |\bA_\perp|^2$& $|A_\perp|^2 - |\bA_\perp|^2$ \\
 4 & ${\rm Re}[A_\|(t)A^*_0(t)]$&${\rm Re}[A_\|A^*_0+\bA_\|\bA^*_0]$&${\rm Re}[A_\|A^*_0-\bA_\|\bA^*_0]$ \\
 5 & ${\rm Im}[A_\perp(t) A^*_0(t)]$&${\rm Im}[A_\perp A^*_0-\bA_\perp \bA^*_0]$&${\rm Im}[A_\perp A^*_0+\bA_\perp \bA^*_0]$ \\
 6 & ${\rm Im}[A_\perp(t) A^*_\|(t)]$&${\rm Im}[A_\perp A^*_\|-\bA_\perp \bA^*_\|]$&${\rm Im}[A_\perp A^*_\|+\bA_\perp \bA^*_\|]$ \\ \hline
 7 & $|A^{(VS)}_+(t)|^2$ & $|A^{(VS)}_+|^2 + |\bA^{(VS)}_+|^2$& $|A^{(VS)}_+|^2 - |\bA^{(VS)}_+|^2$  \\
 8 & ${\rm Re}[A^{(VS)}_+(t) A^*_S(t)]$&${\rm Re}[A^{(VS)}_+A^*_S-\bA^{(VS)}_+\bA^*_S]$&${\rm Re}[A^{(VS)}_+A^*_S+\bA^{(VS)}_+\bA^*_S]$\\
 9 & ${\rm Re}[A^{(VS)}_+(t)A^{(VS)*}_-(t)]$ & ${\rm Re}[A^{(VS)}_+A^{(VS)*}_- - \bA^{(VS)}_+\bA^{(VS)*}_-]$ & ${\rm Re}[A^{(VS)}_+A^{(VS)*}_-+\bA^{(VS)}_+\bA^{(VS)*}_-]$\\
10 & ${\rm Re}[A^{(VS)}_+(t)A^*_0(t)]$&${\rm Re}[A^{(VS)}_+A^*_0-\bA^{(VS)}_+\bA^*_0]$&${\rm Re}[A^{(VS)}_+A^*_0+\bA^{(VS)}_+\bA^*_0]$ \\
11 & ${\rm Re}[A^{(VS)}_+(t)A^*_\|(t)]$&${\rm Re}[A^{(VS)}_+A^*_\|-\bA^{(VS)}_+\bA^*_\|]$&${\rm Re}[A^{(VS)}_+A^*_\|+\bA^{(VS)}_+\bA^*_\|]$ \\
12 & ${\rm Im}[A_\perp(t) A^{(VS)*}_+(t)]$&${\rm Im}[A_\perp A^{(VS)*}_++\bA_\perp\bA^{(VS)*}_+]$&${\rm Im}[A_\perp A^{(VS)*}_+-\bA_\perp\bA^{(VS)}_+]$\\ \hline
13 & $|A^{(VS)}_-(t)|^2$ & $|A^{(VS)}_-|^2 + |\bA^{(VS)}_-|^2$& $|A^{(VS)}_-|^2 - |\bA^{(VS)}_-|^2$ \\
14 & ${\rm Re}[A^{(VS)}_-(t) A^*_S(t)]$&${\rm Re}[A^{(VS)}_-A^*_S+\bA^{(VS)}_-\bA^*_S]$&${\rm Re}[A^{(VS)}_-A^*_S-\bA^{(VS)}_-\bA^*_S]$\\
15 & ${\rm Re}[A^{(VS)}_-(t) A^*_0(t)]$&${\rm Re}[A^{(VS)}_-A^*_0+\bA^{(VS)}_-\bA^*_0]$&${\rm Re}[A^{(VS)}_-A^*_0-\bA^{(VS)}_-\bA^*_0]$\\
16 & ${\rm Re}[A^{(VS)}_-(t) A^*_\|(t)]$&${\rm Re}[A^{(VS)}_-A^*_\|+\bA^{(VS)}_-\bA^*_\|]$&${\rm Re}[A^{(VS)}_-A^*_\|-\bA^{(VS)}_-\bA^*_\|]$\\
17 & ${\rm Im}[A_\perp(t) A^{(VS)*}_-(t)]$&${\rm Im}[A_\perp A^{(VS)*}_--\bA_\perp\bA^{(VS)}_-]$&${\rm Im}[A_\perp A^{(VS)*}_-+\bA_\perp\bA^{(VS)*}_-]$\\ \hline
18 & ${\rm Re}[A_S(t) A^*_\|(t)]$&${\rm Re}[A_SA^*_\|+\bA_S\bA^*_\|]$&${\rm Re}[A_SA^*_\|-\bA_S\bA^*_\|]$\\
19 & ${\rm Im}[A_\perp(t) A^*_S(t)]$&${\rm Im}[A_\perp A^*_S-\bA_\perp\bA^*_S]$&${\rm Im}[A_\perp A^*_S+\bA_\perp\bA^*_S]$ \\
20 & ${\rm Re}[A_S(t)A^*_0(t)]$&${\rm Re}[A_SA^*_0 + \bA_S\bA^*_0]$ & ${\rm Re}[A_S A^*_0 - \bA_S\bA^*_0]$ \\
21 & $|A_S(t)|^2$ & $|A_S|^2 + |\bA_S|^2$& $|A_S|^2 - |\bA_S|^2$ \\ \hline\hline
\end{tabular}
\end{table}

\begin{table}[!htb]
\caption{$b_n$'s and $d_n$'s as defined in Eq.\ (\ref{eq:gentd})
\label{tab:bds}}
\centering
\begin{tabular}{|c|c|c|c|} \hline \hline
 $n$ & $K_n(t)$ & $b_n$ & $d_n$ \\ \hline
 1 & $|A_0(t)|^2$ &$-2~{\rm Re}[A^*_0\bA_0~e^{-\i\phi_M}]$&$-2~{\rm Im}[A^*_0\bA_0~e^{-\i\phi_M}]$ \\
 2 & $|A_\|(t)|^2$ &$-2~{\rm Re}[A^*_\|\bA_\|~e^{-\i\phi_M}]$&$-2~{\rm Im}[A^*_\|\bA_\|~e^{-\i\phi_M}]$\\
 3 & $|A_\perp(t)|^2$&$~2~{\rm Re}[A^*_\perp\bA_\perp~e^{-\i\phi_M}]$&$~2~{\rm Im}[A^*_\perp\bA_\perp~e^{-\i\phi_M}]$ \\
 4 & ${\rm Re}[A_\|(t)A^*_0(t)]$&$-{\rm Re}[(\bA_\|A^*_0+A^*_\|\bA_0)~e^{-\i\phi_M}]$&$-{\rm Im}[(\bA_\|A^*_0+A^*_\|\bA_0)~e^{-\i\phi_M}]$ \\
 5 & ${\rm Im}[A_\perp(t) A^*_0(t)]$&${\rm Im}[(\bA_\perp A^*_0+A^*_\perp\bA_0)~e^{-\i\phi_M}]$&
 $-{\rm Re}[(\bA_\perp A^*_0+A^*_\perp\bA_0)~e^{-\i\phi_M}]$\\
 6 & ${\rm Im}[A_\perp(t) A^*_\|(t)]$&${\rm Im}[(\bA_\perp A^*_\|+A^*_\perp\bA_\|)~e^{-\i\phi_M}]$&
 $-{\rm Re}[(\bA_\perp A^*_\|+A^*_\perp\bA_\|)~e^{-\i\phi_M}]$\\ \hline
 7 & $|A^{(VS)}_+(t)|^2$&$2~{\rm Re}[A^{(VS)*}_+\bA^{(VS)}_+~e^{-\i\phi_M}]$&$2~{\rm Im}[A^{(VS)*}_+A^{(VS)}_+~e^{-\i\phi_M}]$\\
 8 & ${\rm Re}[A^{(VS)}_+(t) A^*_S(t)]$&${\rm Re}[(\bA^{(VS)}_+A^*_S-A^{(VS)*}_+\bA_S)~e^{-\i\phi_M}]$&${\rm Im}[(\bA^{(VS)}_+A^*_S-A^{(VS)*}_+\bA_S)~e^{-\i\phi_M}]$\\
 9 & ${\rm Re}[A^{(VS)}_+(t)A^{(VS)*}_-(t)]$&${\rm Re}[(\bA^{(VS)}_+A^{(VS)*}_-~~~~~~~~~~~$&${\rm Im}[(\bA^{(VS)}_+A^{(VS)*}_+~~~~~~~~~~~$ \\
   & &$~~~~~~~~~~~ -A^{(VS)*}_+\bA^{(VS)}_-)~e^{-\i\phi_M}]$&$~~~~~~~~~~~-A^{(VS)}_+\bA^{(VS)}_+)~e^{-\i\phi_M}]$\\
10 & ${\rm Re}[A^{(VS)}_+(t)A^*_0(t)]$&${\rm Re}[(\bA^{(VS)}_+A^*_0-A^{(VS)*}_+\bA_0)~e^{-\i\phi_M}]$&${\rm Im}[(\bA^{(VS)}_+A^*_0-A^{(VS)*}_+\bA_0)~e^{-\i\phi_M}]$\\
11 & ${\rm Re}[A^{(VS)}_+(t)A^*_\|(t)]$&${\rm Re}[(\bA^{(VS)}_+A^*_\|-A^{(VS)*}_+\bA_\|)~e^{-\i\phi_M}]$&
${\rm Im}[(\bA^{(VS)}_+A^*_\|-A^{(VS)*}_+\bA_\|)~e^{-\i\phi_M}]$\\
12 & ${\rm Im}[A_\perp(t) A^{(VS)*}_+(t)]$&${\rm Im}[(\bA_\perp A^{(VS)*}_+-A^*_\perp\bA^{(VS)}_+)~e^{-\i\phi_M}]$&
$-{\rm Re}[(\bA_\perp A^{(VS)*}_+-A^*_\perp\bA^{(VS)}_+)~e^{-\i\phi_M}]$\\ \hline
13 & $|A^{(VS)}_-(t)|^2$&$-2~{\rm Re}[A^{(VS)*}_-\bA^{(VS)}_-~e^{-\i\phi_M}]$&$-2~{\rm Im}[A^{(VS)*}_-\bA^{(VS)}_-~e^{-\i\phi_M}]$\\
14 & ${\rm Re}[A^{(VS)}_-(t) A^*_S(t)]$&$-{\rm Re}[(\bA^{(VS)}_-A^*_S+A^{(VS)*}_-\bA_S)~e^{-\i\phi_M}]$&$-{\rm Im}[(\bA^{(VS)}_-A^*_S+A^{(VS)*}_-\bA_S)~e^{-\i\phi_M}]$\\
15 & ${\rm Re}[A^{(VS)}_-(t) A^*_0(t)]$&$-{\rm Re}[(\bA^{(VS)}_-A^*_0+A^{(VS)}_-\bA_0)~e^{-\i\phi_M}]$&$-{\rm Im}[(\bA^{(VS)}_-A^*_0+A^{(VS)*}_-\bA_0)~e^{-\i\phi_M}]$\\
16 & ${\rm Re}[A^{(VS)}_-(t) A^*_\|(t)]$&$-{\rm Re}[(\bA^{(VS)}_-A^*_\|+A^{(VS)*}_-\bA_\|)~e^{-\i\phi_M}]$&
$-{\rm Im}[(\bA^{(VS)}_-A^*_\|+A^{(VS)*}_-\bA_\|)~e^{-\i\phi_M}]$\\
17 & ${\rm Im}[A_\perp(t) A^{(VS)*}_-(t)]$&${\rm Im}[(\bA_\perp A^{(VS)*}_-+A^*_\perp\bA^{(VS)}_-)~e^{-\i\phi_M}]$&
$-{\rm Re}[(\bA_\perp A^{(VS)*}_-+A^*_\perp\bA^{(VS)}_-)~e^{-\i\phi_M}]$\\ \hline
18 & ${\rm Re}[A_S(t) A^*_\|(t)]$&$-{\rm Re}[(\bA_SA^*_\|+A^*_S\bA_\|)~e^{-\i\phi_M}]$&$-{\rm Im}[(\bA_SA^*_\|+A^*_S\bA_\|)~e^{-\i\phi_M}]$\\
19 & ${\rm Im}[A_\perp(t) A^*_S(t)]$&${\rm Im}[(\bA_\perp A^*_S+A^*_\perp\bA_S)~e^{-\i\phi_M}]$&
$-{\rm Re}[(\bA_\perp A^*_S+A^*_\perp\bA_S)~e^{-\i\phi_M}]$\\
20 & ${\rm Re}[A_S(t)A^*_0(t)]$&$-{\rm Re}[(\bA_SA^*_0+A^*_S\bA_0)~e^{-\i\phi_M}]$&$-{\rm Im}[(\bA_SA^*_0+A^*_S\bA_0)~e^{-\i\phi_M}]$\\
21 & $|A_S(t)|^2$&$-2~{\rm Re}[A^*_S\bA_S~e^{-\i\phi_M}]$&$-2~{\rm Im}[A^*_S\bA_S~e^{-\i\phi_M}]$\\ \hline\hline
\end{tabular}
\end{table}

\section{Searching for New Physics}

The result of the previous section is quite theoretical. In order to
understand better how to use it to search for NP\footnote{A different
  method for searching for NP in $\btos$ $\bs \to V_1 V_2$ penguin
  decays is discussed in Ref.~\cite{GF}. Here the idea is to use
  flavor SU(3) symmetry and obtain information from measurements of
  the SU(3)-related $\btod$ $B^0$ decays.}, it is necessary to know
what the SM predictions are. Above we focused on final states to which
both $\bs$ and $\bsbar$ can decay. This restricts the analysis to
$\btos$ transitions, so that the quark content of the final state is
$s{\bar s}s{\bar s}$ ($\phi\phi$), $s{\bar s}d{\bar d}$ ($K^{*0} {\bar
  K}^{*0}$), or $s{\bar s}u{\bar u}$ ($K^{*+} K^{*-}$,
$\phi\rho$). Decays to $\phi\phi$ and $K^{*0} {\bar K}^{*0}$ are pure
gluonic penguin decays, and $K^{*+} K^{*-}$ is dominated by the
gluonic penguin (there is a small tree contribution). The decay to
$\phi\rho$ has no gluonic penguin component -- it arises due to
electroweak penguin and tree diagrams. As such, its branching ratio is
quite a bit smaller than that of the other decays, so that even if it
is eventually measured, it is not clear if an angular analysis can be
done.

One quantity that is of interest in such decays is the indirect
(mixing-induced) CP-violating asymmetry (CPA). For a given helicity
$h$, the indirect CPA measures
\beq
\label{indirectCPA}
{\rm Im}\left( e^{-i \phi_M} A_h^* {\bar A}_h \right) ~.
\eeq
Note that the above quantity, which corresponds to the $d_n$ of Table
\ref{tab:bds}, is sensitive to the weak phases of both the mixing and
the decay.  Now, in the SM the gluonic penguin arises dominantly from
the top loop at short distance. Given that $\bs$-$\bsbar$ mixing is
also dominated by the box diagram with an internal top quark, it is
clear that the $e^{-i \phi_M}$ term in Eq.~(\ref{indirectCPA}) cancels
the weak phase in $A_h^* {\bar A}_h$, so that the indirect CPA
vanishes. This is a common argument.

However, things are a bit more complicated. In particular, one also
has to consider the new up and charm penguin amplitudes that are
generated at the $b$ mass scale. These contributions can arise from
the tree-level operators $\btoccs$ and $\btouus$ that produce the
final-state particles through rescattering: ${\bar c} c \rightarrow
\bar{q} q$ and $ {\bar u} u \rightarrow \bar{q} q$ where $q=d,s$. For
a given helicity $h$, the $\btos$ gluonic penguin amplitude can be
written
\bea
\label{amp}
A_h &=& V_{tb}^* V_{ts} P'_{t,h}  + V_{cb}^* V_{cs} P'_{c,h} + V_{ub}^* V_{us} P'_{u,h} \nn\\
    &=& |V_{tb}^* V_{ts}| e^{-i\phi_M/2} P'_{tc,h} + |V_{ub}^* V_{us}| e^{i\gamma} P'_{uc,h} ~.
\eea
(As this is a $\btos$ transition, the diagrams are written with
primes.) In the second line, we have used the unitarity of the
Cabibbo-Kobayashi-Maskawa (CKM) matrix ($V_{tb}^* V_{ts} + V_{cb}^*
V_{cs} + V_{ub}^* V_{us} = 0$) to eliminate the $c$-quark
contribution: $P'_{tc,h} \equiv P'_{t,h} - P'_{c,h}$, $P'_{uc} \equiv
P'_{u,h} - P'_{c,h}$. We have also explicitly written the weak-phase
dependence, while $P'_{tc,h}$ and $P'_{uc,h}$ contain strong phases.

Now, we know that $|V_{tb}^* V_{ts}|$ and $|V_{ub}^* V_{us}|$ are
$O(\lambda^2)$ and $O(\lambda^4)$, respectively, where $\lambda=0.22$
is the sine of the Cabibbo angle. This implies that the $|V_{ub}^*
V_{us}| P'_{uc,h}$ term is much smaller in magnitude than $|V_{tb}^*
V_{ts}| P'_{tc,h}$.
If $|V_{ub}^* V_{us}| P'_{uc,h}$ is neglected, then the $e^{-i
  \phi_M}$ term in Eq.~(\ref{indirectCPA}) cancels the weak phase in
$A_h^* {\bar A}_h$, so that the indirect CPA vanishes. However, while
the result (a vanishing indirect CPA) is correct, the argument leading
to it is not. The easiest way to see this is to use CKM unitarity to
eliminate the $t$-quark contribution in the first line of
Eq.~(\ref{amp}). The amplitude now reads
\beq
A_h = |V_{cb}^* V_{cs}| P'_{ct,h} + |V_{ub}^* V_{us}| e^{i\gamma} P'_{ut,h} ~,
\eeq
where $P'_{ct,h} \equiv P'_{c,h} - P'_{t,h}$, $P'_{ut} \equiv P'_{u,h}
- P'_{t,h}$. Now if $|V_{ub}^* V_{us}| P'_{ut,h}$ is neglected, there
is no cancellation of the $e^{-i \phi_M}$ term in
Eq.~(\ref{indirectCPA}), and the indirect CPA is (apparently)
nonzero. So there appears to be a contradiction.

What is really going on is the following. $|V_{ub}^* V_{us}|$ is
$O(\lambda^4)$. If it is neglected, for consistency one must neglect
{\it all} $O(\lambda^4)$ terms. One of these is ${\rm Im}(V_{tb}^*
V_{ts})$, so that $V_{tb}^* V_{ts}$ is real.  And since $\phi_M
\propto \arg(V_{tb}^* V_{ts})$, it too vanishes in the limit that
$O(\lambda^4)$ terms are neglected.  So, at the end of the day, we
recover the result of a vanishing indirect CPA. The difference is that
here the up and charm penguin contributions have been properly taken
into account.

The weak phase of $A_h$ is therefore generated by keeping the
$|V_{ub}^* V_{us}|$ term. The amplitude can then be written
\bea
A_h &=& |V_{tb}^* V_{ts}| e^{-i\phi_M/2} P'_{tc,h} + |V_{ub}^* V_{us}| e^{i\gamma} P'_{uc,h} \nn\\
    &=& e^{-i\phi_M/2} \left[ |V_{tb}^* V_{ts}| P'_{tc,h} + |V_{ub}^* V_{us}| e^{i(\gamma + \phi_M/2)} P'_{uc,h} \right] ~.
\eea
Writing the strong phases explicitly, we have
\beq
A_h = e^{-i\phi_M/2} \left[ P'_{tc,h} e^{i\delta_{tc,h}} + P'_{uc,h} e^{i(\gamma + \phi_M/2)} e^{i\delta_{uc,h}} \right] ~.
\label{Bsppamp2}
\eeq
In the above, $P'_{tc,h}$ and $P'_{uc,h}$ have been redefined to
absorb $|V_{tb}^* V_{ts}|$ and $|V_{ub}^* V_{us}|$, respectively, so
that $R_h \equiv P'_{uc,h}/P'_{tc,h} = O(\lambda^2)$.  

The SM $a_n$-$d_n$ can be calculated using Tables \ref{tab:acs} and
\ref{tab:bds} with the above expression for $A_h$. These take the
general form $P'_{tc,h} P'_{tc,h'}$ multiplied by either quantities of
$O(1)$ or $R_h \equiv P'_{uc,h}/P'_{tc,h}$. In the SM, those of the
second type are expected to be smaller than those of the first type
since $R_h = O(\lambda^2)$. In fact, the coefficients proportional to
$R_h$ are all CP-violating observables: direct CP asymmetries,
indirect CP asymmetries, triple products, and mixing-induced triple
products.  Physically, this makes sense -- CP violation is due to the
interference of two amplitudes. But in the SM one of the amplitudes
($P'_{uc,h}$) is quite small, so that all CP-violating observables are
also small. As shown below, this is a key point in the search for new
physics.

New physics can enter in two different places -- in $\bs$-$\bsbar$
mixing or in the decay. We discuss these in turn below.

\subsection{NP in the mixing}

If there is NP in $\bs$-$\bsbar$ mixing, this has two consequences.
First, $\phi_M$, which is predicted to be $\simeq 0$ in the SM, could
be large. Second, the weak phase associated with $P'_{tc,h}$ will not,
in general, be equal to $e^{-i\phi_M/2}$, so a nonzero indirect CPA
could appear.  However, LHCb has already measured the phase of
$\bs$-$\bsbar$ mixing in $\Bsdecay$ \cite{LHCbBsmixing}.  They find
\beq
\phi_s = 0.07 \pm 0.09~({\rm stat}) \pm 0.01~({\rm
  syst})~{\rm rad} ~,
\label{2betasmeas}
\eeq
in agreement with the SM. While the errors are large enough that NP
cannot be excluded, a very large deviation from 0 is ruled out.  

Given this, it appears that, at present, $\btos$ penguin decays are
not sensitive to NP in $\bs$-$\bsbar$ mixing. Put another way, it is
probably best to search for such NP using the decay mode $\Bsdecay$.

\subsection{NP in the decay}

The second, more interesting possibility is that there is NP in the
decay. In this case the amplitude takes the form (we neglect
$P'_{uc,h}$ and $\phi_M$)
\beq
A_h = P'_{tc,h} e^{i\delta_{tc,h}} + P'_{NP,h} e^{i\phi_{NP}} e^{i\delta_{NP,h}} ~.
\label{BsppampNP}
\eeq
This has the same form as Eq.~(\ref{Bsppamp2}), with $P'_{uc,h} \to
P'_{NP,h}$, $(\gamma + \phi_M/2) \to \phi_{NP}$ and $\delta_{uc,h} \to
\delta_{NP,h}$. As a result, the expressions for the $a_n$-$d_n$ are
the same as in the SM, with these substitutions.

The signal for NP in the decay is then evident. In the presence of NP,
$R_h$ is equal to $P'_{NP,h}/P'_{tc,h}$, which can be significantly
larger than its SM value, $O(\lambda^2)$. As noted above, the
$a_n$-$d_n$ proportional to $R_h$ all correspond to CP-violating
observables. These can only be large in the presence of a sizeable
second amplitude, i.e., NP in the decay.

\subsection{Measuring CP-violating observables}
\label{measobs}

The bottom line is that the angular analysis of $\bs \to V_1/S_1 (\to
P_1P'_1) V_2/S_2 (\to P_2P'_2)$ is sensitive to NP in the decay.  In
order to search for this NP, measurements must be made of the
CP-violating observables. Here we discuss these in more detail,
referring to Tables \ref{tab:acs} and \ref{tab:bds}. 

As noted above, there are four such observables. In general, the
direct CP asymmetries take the form ${\rm Re}[A_h
  A^*_{h'}-\bA_h\bA^*_{h'}]$ (for $h=h'$, this becomes the familiar
$|A_h|^2 - |\bA_h|^2$). The indirect (mixing-induced) CP asymmetries
are ${\rm Im}[(A^*_h\bA_{h'} + \bA_h A^*_{h'})~e^{-\i\phi_M}]$.

The triple products (TPs) \cite{Datta:2003mj} are a little more
complicated. For a $\bs$ decay, the TP takes the form ${\rm
  Im}[A_\perp A^*_h]$. Now, the amplitudes possess both weak and
strong phases. However, the TP can be nonzero even in the absence of
any weak phases, as long as the strong-phase difference is
nonzero. Thus, a nonzero TP is not necessarily a signal of CP
violation.  In order to obtain a true CP-violating signal, one has to
compare the TPs in $\bs$ and $\bsbar$ decays. This latter TP is given
by ${\rm Im}[{\bar A}_\perp {\bar A}^*_h]$. One combination of $\bs$
and $\bsbar$ TPs is nonzero only if the weak phases are nonzero, and
so is called a true (CP-violating) TP. The second combination can be
nonzero even if the weak phases are zero, and so it is not a signal of
CP violation -- it is called a fake TP. The true TP takes the form
${\rm Im}[A_\perp A^*_h - \bA_\perp \bA^*_h]$. One can also have TPs
induced by $\bs$-$\bsbar$ mixing. The true mixing-induced TP is ${\rm
  Im}[(\bA_\perp A^*_h+A^*_\perp\bA_h)~e^{-\i\phi_M}]$.

The coefficients corresponding to the four CP-violating observables
are
\begin{enumerate}

\item direct CP asymmetries: $c_n$
  ($n=1{\hbox{-}}4,7,13{\hbox{-}}16,18,20,21$), $a_n$
  ($n=8{\hbox{-}}11$);

\item indirect CP asymmetries: $d_n$
  ($n=1{\hbox{-}}4,7,13{\hbox{-}}16,18,20,21$), $b_n$
  ($n=8{\hbox{-}}11$);

\item triple products: $a_n$ ($n=5,6,17,19$), $c_{12}$;

\item mixing-induced triple products: $b_n$ ($n=5,6,17,19$), $d_{12}$.

\end{enumerate}
If any coefficient is measured to be significantly larger than the SM
prediction ($O(\lambda^2)$), this would be a sign of NP in the decay.
Note that, in general, the direct CP asymmetries, indirect CP
asymmetries, TPs, and mixing-induced TPs are represented by the $c_n$,
$d_n$, $a_n$ and $b_n$, respectively. However, this pattern is broken
for $n=8{\hbox{-}}12$. The reason is that these observables involve
$A^{(VS)}_+$, which is CP odd. That is, in going from the $\bs$ to
$\bsbar$ decay, $A^{(VS)}_+ \to - \bA^{(VS)}_+$. This additional minus
sign leads to the pattern breaking above.

As noted in Sec.~\ref{sec:tiad}, it is {\it not} necessary to do the full
angular analysis to measure these observables. Rather, by performing
asymmetric integrals over the three angles, one can isolate (almost)
any angular function, i.e., value of $n$.  Then one uses the
time-dependence of Eq.~(\ref{eq:gentd}) to distinguish among the
$a_n$-$d_n$. Indeed, this has already been done in
Refs.~\cite{CDFBs,LHCbphiphi} for the TPs $a_5$ and $a_6$ in the
simpler case of time-integrated untagged $\bs \to \phi \phi$ decays.

In fact, it should be stressed that at this stage there is no point in
trying to perform a full angular analysis. The aim of such an analysis
would be to determine the NP parameters.  However, unless a NP signal
is found, this is irrelevant. We therefore suggest that experiments
concentrate on measuring the $a_n$-$d_n$ that are expected to be small
in the SM.

In the following two sections we discuss the above formalism in the
context of the specific decays $\bs \to \phi \phi$ and $\bs \to K^{*0}
{\bar K}^{*0}$.

\section{\boldmath $\bs \to \phi \phi$}

The LHCb Collaboration studied $\bs \to \phi \phi$ in
Ref.~\cite{BsphiphiLHCb}. Their analysis uses the following logic.
They argue that, since $\bs \to \phi \phi$ proceeds via a gluonic
${\bar b} \to {\bar s}s{\bar s}$ diagram with a $t$ quark in the loop,
the mixing and decay weak phases cancel identically. QCD factorization
calculations, which take into account the up and charm penguin
contributions, find an upper limit of 0.02 rad for $|\theta_s|$
\cite{QCDf}, where $\theta_s$ is the phase in the $\bs \to \phi \phi$
decay. Putting this all together, LHCb writes each helicity amplitude
as
\beq
A_h = |A_h| e^{i\theta_s/2} e^{i\delta_h} ~,
\label{Bsppamp}
\eeq
where $\delta_h$ is the strong phase. Here $\theta_s$ is taken to be a
weak phase and is assumed to be helicity-independent.

The angular analysis of $\bs \to \phi \phi$ is as given in Sec.~2,
except that, for this final state, we have $|VS\rangle = -|SV\rangle$
so that the $VS_+$ state vanishes. This implies that the
$A^{(VS)}_+(t)$ amplitude is not present. The expressions for the
coefficients $a_n$-$d_n$ are found using the formulae in Tables
\ref{tab:acs} and \ref{tab:bds}, in which $A_h$ is given in
Eq.~(\ref{Bsppamp}). These are listed in Table \ref{tab:Bspp1} for
$n=1$-6 (in total, $n$ goes to 15, but LHCb finds that the $VS$ and
$SS$ contributions are very small). We define $\delta_1 \equiv
\delta_\perp - \delta_\|$, $\delta_2 \equiv \delta_\perp - \delta_0$,
and $\delta_{2,1} \equiv \delta_2 - \delta_1$.  Table \ref{tab:Bspp1}
agrees with Ref.~\cite{BsphiphiLHCb}.

\begin{table}[htb]
\caption{$a_n$-$d_n$'s ($n=1$-6) for $\bs \to \phi \phi$. $A_h$ takes
  the form in Eq.~(\ref{Bsppamp}).
\label{tab:Bspp1}}
\centering
\begin{tabular}{|c|c|c|c|c|c|} \hline \hline
 $n$ & $N$ & $a_n/N$ & $b_n/N$ & $c_n/N$ & $d_n/N$ \\ \hline
 1 & $2|A_0|^2$ & 1 & $-\cos \theta_s$ & 0 & $\sin \theta_s$ \\
 2 & $2|A_\||^2$ & 1 & $-\cos \theta_s$ & 0 & $\sin \theta_s$ \\
 3 & $2|A_\perp|^2$ & 1 & $\cos \theta_s$ & 0 & $-\sin \theta_s$ \\
 4 & $2|A_0||A_\||$ & $\cos\delta_{2,1}$ & $- \cos\delta_{2,1} \cos \theta_s$ & 0 & $ \cos\delta_{2,1} \sin \theta_s$ \\
 5 & $2|A_0||A_\perp|$ & 0 & $- \cos\delta_2 \sin \theta_s$ & $ \sin\delta_2$ & $- \cos\delta_2 \cos \theta_s$ \\
 6 & $2|A_\|||A_\perp|$ & 0 & $- \cos\delta_1 \sin \theta_s$ & $\sin\delta_1$ & $- \cos\delta_1 \cos \theta_s$ \\ 
\hline\hline
\end{tabular}
\end{table}

Of course, the form assumed for the $A_h$ [Eq.~(\ref{Bsppamp})] has
specific implications for the expressions for the $a_n$-$d_n$. For
example, the fact that $c_n = 0$ ($n=1$-4) and $a_n = 0$ ($n=5$,6) is
a direct consequence. In addition, the quantity $\sin \theta_s$
appears explicitly in a number of entries in Table \ref{tab:Bspp1}, so
that $\sin \theta_s = \pm d_n/a_n$ ($n=1$-4). Also, $\tan \theta_s =
b_n/d_n$ ($n=5,6$).  This allows LHCb to restrict the value of
$\theta_s$ to the interval of $[-2.46,-0.76]$ rad at 68\% C.L. As
explained above, $\theta_s$ is expected to be quite small in the SM,
so this result is an intriguing hint of NP.

The problem is that the logic leading to Eq.~(\ref{Bsppamp}) is
somewhat faulty, and the form assumed for $A_h$ is not the most
general. Indeed, a rather strong assumption has been made, one that is
not well-motivated physically. The exact amplitude is given in
Eq.~(\ref{Bsppamp2}). One can rephase it by $e^{i\phi_M/2}$, giving
\bea
\label{sqbrackets}
A_h & = & P'_{tc,h} e^{i\delta_{tc,h}} + P'_{uc,h} e^{i(\gamma + \phi_M/2)} e^{i\delta_{uc,h}} \nn\\
& = & P'_{tc,h} e^{i\delta_{tc,h}} \left[ 1 + R_h e^{i(\gamma + \phi_M/2)e^{i \Delta_h}} \right] ~,
\eea
where $\Delta_h=\delta_{uc,h}-\delta_{tc,h} $. If we assume that
$\Delta_h = 0$ and that $R_h$ is helicity-independent, we can write
the piece in square brackets as $X e^{i\theta_s/2}$, which yields
Eq.~(\ref{Bsppamp}). Here we see that $\theta_s$ is indeed small,
$O(\lambda^2)$, and this is due to the smallness of $R_h$.

Unfortunately, the assumption that $\Delta_h = 0$ is generally false
and is not supported by any computation. Given that the starting
point, Eq.~(\ref{Bsppamp}), is questionable, this raises serious
warning signs about the result.  Is the large measured value of
$\theta_s$ really a hint of NP, or is it just a consequence of an
incorrect assumption?

Still, one question remains: although Eq.~(\ref{Bsppamp}) is not an
exact parametrization of the amplitude, could it be taken as an
approximation? And here the answer is yes. The key point is that, even
if $\Delta_h = 0$ is not assumed, the piece in square brackets in
Eq.~(\ref{sqbrackets}) can be written as $X e^{i\theta_{s,h}/2}$.
However, as $\theta_{s,h} $ is small in the SM, the assumptions that
it is helicity independent and purely a weak phase are acceptable
provided one is not measuring quantities of the expected size of
$\theta_{s} $, $O(\lambda^2)$. In other words, if a value of
$\theta_{s}$ is measured that is much larger than the SM expectation,
then this would be a sign of NP. However, if the measured $\theta_{s}$
is small, then even if it deviates from the SM expectation, one cannot
reliably claim the presence of NP.

But this then raises another question: if such large NP effects are
present, are they best detected by performing a fit to the data? Here
the answer is clearly no. The only way one can find a large value of
$\theta_{s}$ is if there are CP-violating observables whose values are
much larger than in the SM. But in this case, the simple measurement
of these observables will reveal the presence of NP -- it is not
necessary to perform a full fit to the data.

The bottom line is that if one wishes to perform a fit, it is best to
do the analysis using the exact amplitude of Eq.~(\ref{Bsppamp2}).
This will be difficult, as there are considerably more unknown
parameters than in Eq.~(\ref{Bsppamp}). However, as stated previously,
at this stage a full angular analysis is not even warranted.
Experiments measuring $\bs \to \phi \phi$ should simply focus on
measuring the CP-violating observables, since these are expected to be
small in the SM.

\section{\boldmath $\bs \to K^{*0} {\bar K}^{*0}$}

As noted in the previous section, LHCb finds that the $VS$ and $SS$
contributions to $\bs \to \phi \phi$ are very small. This is not
surprising. The resonant scalar background comes from the decay $f_0
\to K^+ K^-$. However, the dominant $f_0$ decay is to $\pi\pi$ -- the
Particle Data Group notes only that $f_0 \to K{\bar K}$ has been
``seen'' \cite{pdg}. The $f_0 \to \pi\pi$ decay is an important
background for decays in which a final-state $\rho^0$ is
produced. Indeed, measurements of the decay $B^0 \to \rho^0 \rho^0$
\cite{Brho0rho0} had to take this background into account. However,
$\bs$ decays to final states involving a $\rho^0$ are rare. 

One decay for which the scalar-background contributions are clearly
significant is $\bs \to K^{*0} {\bar K}^{*0}$ \cite{Bernardo}. Here the
final-state vector meson is the $K^{*0}(892)$, identified through its
decay to $K^+ \pi^-$. However, the scalar meson $K^{*0}(1430)$ decays
almost exclusively to the same final state. And since its width is
$(270 \pm 80)$ MeV \cite{pdg}, it constitutes an important
background. Finally, since the final state does not involve identical
particles, both additional amplitudes $A_{VS}$ and $A_{SV}$ (or
equivalently $A^{(VS)}_+$ and $A^{(VS)}_-$ [Eq.~(\ref{CPVS})]) must be
considered.

When this angular analysis is done, the first step will be to examine
the untagged decays. In order to find which observables are present in
these decays, one proceeds as follows.  The time-dependent transversity
amplitudes for the $\bsbar$ decay ($\bar{K}_n$'s) can be obtained by
interchanging $A_h \leftrightarrow \eta_h \bar{A}_h$ and changing the
sign of the weak phase $\phi_M$. One can write an equation similar to
Eq.\ (\ref{eq:gentd}) in the $\bsbar$ case as follows:
\bea\label{eq:gentdb}
\bar{K}_n(t) = \frac12 e^{ - \Gamma t} \left[\bar{a}_n \cosh{(\Delta \Gamma/2) t} + \bar{b}_n \sinh{(\Delta\Gamma/2) t}
+ \bar{c}_n \cos{\Delta m t} + \bar{d}_n \sin{\Delta m t} \right]~,
\eea
where once again $\bar{a}_n, \bar{b}_n, \bar{c}_n$, and $\bar{d}_n$
for $n = 1, \dots, 21$ are time-independent functions of $A_h$ and
$\bA_h$.

It is straightforward to show that
\beq
\bar{a}_n = a_n ~~,~~~~ \bar{c}_n = -c_n ~~,~~~~ \bar{b}_n = b_n ~~,~~~~ \bar{d}_n = -d_n ~.
\eeq
With these results the transversity amplitudes for the
\textit{untagged} decay are
\beq
\label{eq:genunt}
K^{\rm untagged}_n(t) = K_n(t) + \bar{K}_n(t)
  = e^{ - \Gamma t} \left[a_n \cosh{(\Delta \Gamma/2) t} + b_n \sinh{(\Delta\Gamma/2) t}\right]~.
\eeq

As stressed above, experiments should focus on measuring the
CP-violating observables. In the untagged case, these are (see
Sec.~\ref{measobs})
\begin{enumerate}

\item triple products: $a_n$ ($n=5,6,17,19$);

\item mixing-induced triple products: $b_n$ ($n=5,6,17,19$);

\item direct CP asymmetries: $a_n$ ($n=8{\hbox{-}}11$);

\item indirect CP asymmetries: $b_n$ ($n=8{\hbox{-}}11$).

\end{enumerate}
As can be seen, all four types of CP-violating observables are
accessible in untagged $\bs \to K^{*0} {\bar K}^{*0}$ decays, given
that the scalar-background contributions are important.

It is perhaps surprising to find direct CP asymmetries in the untagged
sample. After all, we usually think that such observables require
tagging. However, their presence in the above list can be understood
as follows. As mentioned earlier, the general form for a direct CP
asymmetry is ${\rm Re}[A_h A^*_{h'}-\bA_h\bA^*_{h'}]$. Now, the
$B$-decay contribution is ${\rm Re}[A_h A^*_{h'}]$, while that for the
${\bar B}$ is obtained by taking $A_{h,h'} \to \eta_{h,h'}
\bA_{h,h'}$, where $\eta_{h,h'} = +1$ ($-1$) if $A_{h,h'}$ is CP even
(CP odd). The observable in untagged decays is then the sum of the $B$
and ${\bar B}$ contributions:
\beq
{\rm Re}[A_h A^*_{h'} + \eta_h \eta_{h'} \bA_h\bA^*_{h'}] ~.
\label{dirCPA}
\eeq
Suppose first that both $A_h$ and $A_{h'}$ are CP even. In this case
$\eta_h \eta_{h'} = +1$ and the observable in Eq.~(\ref{dirCPA}) is
not a direct CP asymmetry. This is the case for $n=4$, for example. On
the other hand, suppose that one of $A_h$ and $A_{h'}$ is CP odd. Now
we have $\eta_h \eta_{h'} = -1$ and the observable in
Eq.~(\ref{dirCPA}) {\it is} a direct CP asymmetry. This is what is
occurring for $n=8{\hbox{-}}11$. Here the CP-odd amplitude
$A^{(VS)}_+$ is involved, and this leads to the direct CP asymmetries
in the untagged sample. A similar logic applies to the indirect CP
asymmetries.

\section{New Physics}

Above we have stressed that measurements should be made of the
CP-violating observables. Such measurements are sensitive to NP in the
decay. In this section we examine a model of NP that can yield such
effects. Although we focus on a particular NP scenario, our analysis
is easily applicable to other NP models.

The recent discovery of a Higgs-like resonance at the LHC
\cite{ATLASCMS_h}, along with supporting evidence for its existence
from Fermilab \cite{CDFD0_h}, have renewed interest in models with an
extended Higgs sector. An accurate determination of the couplings of
the new state to quarks, including flavor-changing neutral-current
(FCNC) couplings, is clearly very important. Constraints on possible
FCNC couplings of this state have recently been examined
\cite{HiggsFCNC}.

We consider a model with an extended Higgs sector in which the neutral
scalars have FCNC couplings. We identify the lowest-mass state as the
$X$ particle, and assume that $X$ mediates the decay $\btosqq$, where
$q=d,s$. (There may be contributions to the decay from heavier states,
but in order to retain predictive power, we assume that that the
dominant contribution comes from the lowest-lying state $X$.) $X$ may
be identified with the newly-discovered particle of mass $\sim 125 $
GeV, but this is not important for our discussion. However, it should
be pointed out that the penguin $\bs\to V_1V_2$ decays have the
potential to explore the coupling of the new scalar state to light
quarks, which is not possible at collider experiments.

After integrating out the $X$ state, we generate the effective
Hamiltonian \cite{Datta:2004re,Baek:2005jk}
\bea 
\label{NPoperators} 
H_{NP} & = & {4 G_{\sss F} \over \sqrt{2}} \sum_{\sss A,B = L,R}  f_q^{\sss
AB} \, {\bar b} \gamma_{\sss A} s \, {\bar q} \gamma_{\sss B} q   ~,
\eea
with a total of four contributing operators ($A,B = L,R$, $q=d,s$). In
order to determine the contribution of each operator to the various
observables, it is necessary to calculate the hadronic matrix
elements. However, instead of computing these using a particular
model, we prefer to simply make some general observations. To do this,
we introduce two small parameters:
\begin{enumerate}

 \item We assume that the NP contribution to any observable is smaller
   than that of the dominant SM amplitude, but larger than the
   subdominant SM amplitude of $O(\lambda^2)$ . This is reasonable
   since larger NP contributions would likely have already been seen
   in experiments. We therefore define $\epsilon \equiv |NP|/|SM|$,
   and take its value to be $\sim 20 $\%.

\item We introduce the heavy-quark expansion parameter $\epsilon_b
  \sim \Lambda_{QCD}/m_b$. Generically, we expect that $\epsilon_b
  \sim $ 10-20 \%.

\end{enumerate}

We also make the assumption that the NP matrix elements can be
estimated by naive factorization. This is very reasonable since any
correction to naive factorization would typically be $O(\epsilon
\epsilon_b)$. Note that, with this assumption, the scalar operators
cannot directly produce the $VV$ or $VS$ final states -- they can only
do so after a Fierz transformation. Moreover, as the NP operators do
not contain charm quarks, possible large nonperturbative rescattering
effects are absent \cite{MuruKagan}. A consequence of this is that the
NP amplitudes have strong phases that are 0 or $\pi$. (This can be
justified on more general grounds \cite{Datta:2004re, Datta:2004jm}.)
Since we are considering NP effects with large new weak phases, we can
neglect the small weak phases in the SM amplitudes. Hence the SM and
NP amplitudes, respectively $s_h$ and $n_h$, take the following forms:
\beq
s_h = {\bar s}_h = |s_h| e^{ i\delta_h} ~~,~~~~
n_h = {\bar n}_h^*= |n_h|e^{ i\phi_h} ~, 
\label{amp_rel}
\eeq
where $ \phi_h$ are the NP weak phases and $\delta_h$ are the SM
strong phases.

To simplify things we concentrate on one NP operator at a time, and
consider its effect on the process $\BsKK$ (for this decay, a Fierz
transformation is also needed to produce the $SS$ final state). The
procedure is to obtain the form of each helicity amplitude $A_h$ in
the presence of the NP operator, and then to compute the CP-violating
observables that appear in the untagged distribution. We focus on the
triple products $a_n$ ($n=5,6,17,19$) and the direct CP asymmetries
$a_n$ ($n=8{\hbox{-}}11$). Assuming the mixing phase to be small, the
mixing-induced triple products $b_n$ ($n=5,6,17,19 $) do not provide
additional information over that already contained in the triple
products, and so we do not calculate them. Similarly, the indirect CP
asymmetries $b_n$ ($n=8{\hbox{-}}11$) contain the same information as
the $a_n$ ($n=8{\hbox{-}}11$) when the mixing phase is neglected.  The
triple products arise from the interference of $A_{\perp}$ with the
other amplitudes, while the direct CP asymmetries arise from the
interference of $A_{+}^{(VS)}$ with the other amplitudes.  Note that,
while there are SM contributions to all the $A_h$, $s_\| = -
s_{\perp}$ in the heavy-quark limit \cite{Datta:2011qz}.

We consider the following three cases.
\begin{itemize}

\item Case a: We begin with the NP operator of $\btosdd$ whose
  coefficient is $f_d^{\sss RR}$ [Eq.~(\ref{NPoperators})]:
\beq {4 G_{\sss F} \over \sqrt{2}} f_d^{\sss RR} \, {\bar b}
\gamma_{\sss R} s \, {\bar d} \gamma_{\sss R} d ~. \eeq
For this operator to contribute to the decay, we perform a Fierz
transformation (both fermions and colors):
\beq -{4 \over N_c} {G_{\sss F} \over \sqrt{2}} f_d^{\sss RR} \,
\left[
  \frac12 \, {\bar b} \gamma_{\sss R} d \, {\bar d} \gamma_{\sss R} s
  + \frac18 \, {\bar b} \sigma^{\mu\nu} \gamma_{\sss R} d \, {\bar d}
  \sigma_{\mu\nu} \gamma_{\sss R} s \right] ~.
\label{tensor} \eeq
Under the factorization assumption the currents produce the
final-state mesons, so that the scalar currents cannot produce vector
mesons and the tensor currents cannot produce scalar mesons.  Thus the
first term contributes only to the $SS$ state within factorization and
in the heavy-quark limit, while the second term contributes only to
the $VV$ states. We will make use of the following factorization
results.
\begin{itemize}

\item To leading order in $1/m_b$ we have
\beq
\bra {VV}{\bar b} \gamma_{\sss R} d \, {\bar d} \gamma_{\sss R} s \ket B = 0 ~~,~~~~
\bra {(VS)_{\pm}}{\bar b} \gamma_{\sss R} d \, {\bar d} \gamma_{\sss R} s \ket {B} = 0 ~.
\eeq
The results above are due to $ \bra{V}{\bar b} \gamma_{\sss R} d
\ket{B}=0 $ \cite{Charles:1998dr} and $\bra{V} {\bar d} \gamma_{\sss
  R} s\ket{0}=0$.

\item The matrix element $\bra{VV}{\bar b} \sigma^{\mu\nu}
  \gamma_{\sss R} d \, {\bar d} \sigma_{\mu\nu} \gamma_{\sss R} s
  \ket{B} $ was worked out in Refs.~\cite{Baek:2005jk, Datta:2011qz},
  with the result that the contribution to the longitudinal amplitude
  is $ \sim 1/m_b$ while the transverse amplitudes are unsuppressed.
  We also note that for the $VS$ states, the amplitude in which the
  scalar is produced from the vacuum vanishes as tensor operators
  cannot produce a scalar meson from the vacuum.  The amplitude in
  which the vector state is produced from the vacuum can be shown to
  be suppressed by $\sim 1/m_b$. It is also true for reasons stated
  above that the tensor operators cannot produce the $SS$ state.

\end{itemize} 

Using the results discussed above, and keeping terms up to linear in
$\epsilon $ and $\epsilon_b$, we can write the amplitudes as
\bea
A_0 = s_0 & ~,~~ & A_\perp  = s_{\perp} + n_{\perp}^{RR} ~, \nonumber\\
A_{\|} = - s_{\perp} + n_{\perp}^{RR} + O(\epsilon_b) & ~,~~ & A_{+}^{(VS)} =  s_{+}^{(VS)}~, \nonumber\\
A_{-}^{(VS)} = s_{-}^{(VS)} & ~,~~ & A_{SS} = s_{SS} +n_{SS}^{RR} ~.
\label{NPRR}
\eea
The prediction for this operator is that one should observe nonzero
values for all the triple products while the direct CP-violation terms
$a_9$ and $a_{10}$ should be small.

\item Case b: Similar to the example above, for the $f_d^{\sss LL}$
  operator the amplitudes are
\bea
A_0 = s_0 & ~,~~ & A_\perp  = s_{\perp} + n_{\perp}^{LL}~, \nonumber\\
A_{\|} = - s_{\perp} - n_{\perp}^{LL} + O(\epsilon_b) & ~,~~ & A_{+}^{(VS)} =  s_{+}^{(VS)}~, \nonumber\\
A_{-}^{(VS)} = s_{-}^{(VS)} & ~,~~ & A_{SS} = s_{SS} +n_{SS}^{LL} ~.
\eea
Hence the prediction is that all true triple products have similar
sizes, except for $a_6$ which should be small. As in the previous
case, the direct CP-violation terms $a_9$ and $a_{10}$ should be
small.

\item Case c: Finally, we consider the case when we have the
  operators, $f_d^{\sss LR}$ and $f_d^{\sss RL}$
  [Eq.~(\ref{NPoperators})]. The Fierz transformation produces $(V-A)
  \times (V+A)$ and $(V+A) \times (V-A)$ operators.  In this case the
  NP transverse amplitudes are suppressed by $\epsilon_b$. We can
  write the amplitudes as,
\bea
A_0 = s_0+ n_0^{LR}- n_0^{RL} & ~,~~ & A_\perp = s_{\perp} ~,  \nonumber\\
A_{\|} = - s_{\perp} + O(\epsilon_b) & ~,~~ & A_{+}^{(VS)} = s_{+}^{(VS)} +n_+^{LR, VS}- n_+^{RL,VS} ~,\nonumber\\
A_{-}^{(VS)} = s_{-}^{(VS)} +n_-^{LR, VS}- n_-^{RL,VS} & ~,~~ & A_{SS} = s_{SS} +n_{SS}^{LR}- n_{SS}^{RL} ~.
\eea
In this case the triple product $a_6$ is small, but the other TPs
will generally be nonzero. This case is different from Case b above
as the direct CP-violation terms $a_9$ and $a_{10}$ are not small.

\end{itemize}

We therefore see that the three cases make different predictions for
the CP-violating terms in the untagged distribution. As a result, one
can learn about the nature of the underlying NP from the pattern
of the measurements. If the tagged measurements are also available,
then the additional CP-violating observables can be used to further
pinpoint the structure of the NP.

\section{Conclusions}

It is well known that the amplitude for $B\to V_1V_2$ ($V_i$ is a
vector meson) can be decomposed in terms of three helicities -- $A_0$,
$A_\|$, $A_\perp$ -- and that these can be separated experimentally by
performing an angular analysis of the decay. Recently it was pointed
out that if a neutral vector meson is detected via its decay $V \to P
P'$ ($P, P'$ are pseudoscalars), there is usually a background coming
from scalar resonant or non-resonant $P P'$ production. This can be
taken into account by adding another (scalar) helicity to the angular
analysis.

Since the $\phi$ is detected through its decay to $K^+ K^-$, LHCb
performed this addition in their studies of $\bs \to J/\psi \phi$
\cite{BsJpsiphiLHCb} and $\bs \to \phi \phi$ \cite{BsphiphiLHCb}. For
the first decay there were four helicities in the angular analysis,
while in the second there were five. LHCb is also examining $\bs \to
K^{*0}(892) {\bar K}^{*0}(892)$. In this case, the angular analysis
requires six helicities since there are no identical particles in the
final state.

Also, in its analysis of $\bs \to \phi \phi$, LHCb did not use the
most general decay amplitude. This raises the question of whether the
result of the analysis (an intriguing hint of NP) is due to the chosen
form of the amplitude.

In this paper, we address the above issues. We present the most
general (six-helicity) angular analysis of $\bs \to V_1 (\to P_1P'_1)
V_2 (\to P_2P'_2)$. We focus on final states to which both $\bs$ and
$\bsbar$ can decay. These are mainly $\btos$ penguin transitions. We
also derive the most general decay amplitude. We show that the
amplitude used by LHCb in Ref.~\cite{BsphiphiLHCb} makes an assumption
regarding the strong phases that is not reproduced by direct
calculation.

One of the reasons that LHCb used its form of the decay amplitude is
that it contains a small number of unknown parameters. This permits a
search for NP via a full fit to the data. However, the most general
amplitude contains more unknowns, so that a full fit is considerably
more difficult. Fortunately, a fit is not necessary to detect NP.
Since $\btos$ penguin decays are dominated by a single contributing
amplitude in the SM, all CP-violating observables are predicted to be
small. The presence of NP would then be clearly indicated by the
simple measurement of a sizeable CP-violating observable. There are
four such observables -- direct CP asymmetries, indirect CP
asymmetries, triple products, and mixing-induced triple products --
and we discuss all of these in the context of the six-helicity angular
analysis.

We apply this analysis to the decay $\bs \to K^{*0} {\bar K}^{*0}$. In
particular, we examine which CP-violating observables remain in the
untagged data sample. Triple products and mixing-induced triple
products are of course present. In addition, because this decay has a
CP-odd background, certain direct and indirect CP asymmetries can be
observed in untagged decays. This is a particuliarly interesting
aspect of $\bs \to K^{*0} {\bar K}^{*0}$.

Finally, one can learn about the nature of the underlying NP by
determining which of the CP-violating observables are nonzero. To
demonstrate this, we consider a particular model of NP and show that
different NP operators make different predictions for the pattern of
sizeable CP-violating observables.

\bigskip
\noindent
{\bf Acknowledgments}: A special thanks goes to Bernardo Adeva for his
invaluable input to this project.  We also thank Paula \'Alvarez,
Brais Sanmartin, Antonio Romero, Sean Benson, Franz Muheim, St\'ephane
Monteil and Olivier Leroy for helpful communications. This work was
financially supported by NSERC of Canada (BB, DL), and by the National
Science Foundation under Grant No.\ NSF PHY-1068052 (AD, MD).

\end{document}